\documentclass[pdflatex,sn-standardnature]{sn-jnl}

\jyear{2022}%

\theoremstyle{thmstyleone}%
\theoremstyle{thmstyletwo}%

\theoremstyle{thmstylethree}%

\DeclareGraphicsExtensions{.eps,.eps.gz,.epsi, .jpg, .png}

\newcommand{\teff}{\ensuremath{T_{\mathrm {eff}}\,}}
\newcommand{\logg}{\ensuremath{{\mathrm {log}\, } g\,}}
\newcommand{\feh}{\ensuremath{[{\mathrm {Fe/H}}]\,}}
\newcommand{\afe}{\ensuremath{[{\mathrm {\alpha/Fe}}]\,}}
\newcommand{\mk}{\ensuremath{\mathrm M_K\,}}

\begin{document}

\title[Article Title]{\bf A time-resolved picture of our Milky Way's early formation history}
\author*[1]{\fnm{Maosheng} \sur{Xiang}}\email{mxiang@mpia.de}
\author*[1]{\fnm{Hans-Walter} \sur{Rix}}\email{rix@mpia.de}
\affil[1]{\orgdiv{}, \orgname{Max-Planck Institute for Astronomy}, \orgaddress{\street{K\"onigstuhl 17}, \city{Heidelberg}, \postcode{D-69117}, 
\country{Germany}}}
\maketitle

{\bf The formation of our Milky Way can be parsed qualitatively into different phases that resulted in its structurally different stellar populations: the halo and the disk components \cite{Xiang2015, Bland-Hawthorn-Gerhard2016, Spitoni2019}. Revealing a quantitative overall picture of our Galaxy’s assembly awaits a large sample of stars with very precise ages. Here we report an analysis of such a sample using subgiant stars. We find that the stellar age-metallicity distribution $p(\tau, \feh)$ splits into two almost disjoint parts, separated at $\tau\simeq8$~Gyr. The younger reflecting a late phase of quiescent Galactic disk formation with manifest evidence for stellar radial orbit migration \cite{Frankel2018,Feuillet2019, Wu2021}; the other reflecting the earlier phase, when the stellar halo \citep{Helmi2020} and the old $\alpha$-process-enhanced (`thick') disk \citep{Hayden2015, Bonaca2020} formed. Our results indicate that the formation of the Galaxy’s old (thick) disk started $\sim13$\,Gyr ago, only 0.8\,Gyr after the Big Bang, and two Gigayears earlier than the final assembly of the inner Galactic halo. Most of these stars formed $\sim$11~Gyr ago, when the Gaia-Sausage-Enceladus satellite merged with our Galaxy \cite{Belokurov2018, Helmi2018}. Over the next 5--6~Gyr, the Galaxy experienced continuous chemical element enrichment, ultimately by a factor of 10, while the star-forming gas managed to stay well-mixed.}

To unravel the assembly history of our Galaxy we need to learn how many stars were born when, from what material, and on what orbits. This requires precise age determinations for large sample of stars that extend to the oldest possible ages ($\sim 14$~Gyrs) \cite{Xiang2017, Bonaca2020}. Subgiant stars, stars sustained by hydrogen shell fusion, can be unique tracers for such purposes, as they are in the brief stellar evolutionary phase that permits the most precise and direct age determination, because their luminosity is a direct measure of their age.  Moreover, the chemical element compositions determined from spectra of their photosphere surfaces, accurately reflect their birth material composition billions of years ago. This makes subgiants the best practical tracers of Galactic archaeology, even compared to main-sequence turn-off stars whose surface abundances may be altered by atomic diffusion effects \cite{Dotter2017}. However, due to the short lifetime of their evolutionary phase, subgiant stars are relatively rare, and large surveys are essential to build a large sample of these objects with good spectra, which have not been available in the past.

With the recent data release (eDR3) of the Gaia mission \cite{Prusti2016, Brown2021} and the recent data release (DR7) of the LAMOST spectroscopic survey \citep{Cui2012, Zhao2012}, we identify a set of $\sim250,000$ subgiant stars based on their position in the \teff-\mk diagram (\textbf{Figure~1}). The ages ($\tau$) of these subgiant stars are estimated by fitting to the Yonsei-Yale (YY) stellar isochrones \citep{Demarque2004} with a Bayesian approach, that draws on the astrometric distances (parallaxes), apparent magnitudes (fluxes), spectroscopic chemical abundances (\feh, [$\alpha$/Fe]), effective temperatures (\teff), and absolute magnitudes \mk. As summarized in \textbf{Figure~1}, the sample stars have a median relative age uncertainty of only 7.5\% across the age range from 1.5\,Gyr to the age of the universe \cite[13.8~Gyr;][]{Planck2016}. The lower age limit of our sample is inherent to our approach: younger and hence more luminous subgiants can be confused with a different stellar evolutionary phase, the horizontal branch phase in far older stars, which would cause serious sample contamination. This sample constitutes a 100-fold leap in sample size for stars with comparably precise and consistent age estimates \citep{SilvaAguirre2017, Montalban2021}. And it is the first large sample to cover a large spatial volume across the Milky Way ($6\lesssim R\lesssim14$\,kpc, $-4<Z<6$\,kpc) and most of the pertinent range in age and in metallicity, \feh : $1.5<\tau<13.8$\,Gyr and $-2.5<\feh < 0.4$. The sample also has a straightforward spatial selection function that allows us to estimate space density of the tracers. These ingredients enable a new view of the Milky Way's assembly history, especially the early formation history. \\

\begin{figure}[htp!]
\centering
\includegraphics[width=1.0\textwidth]{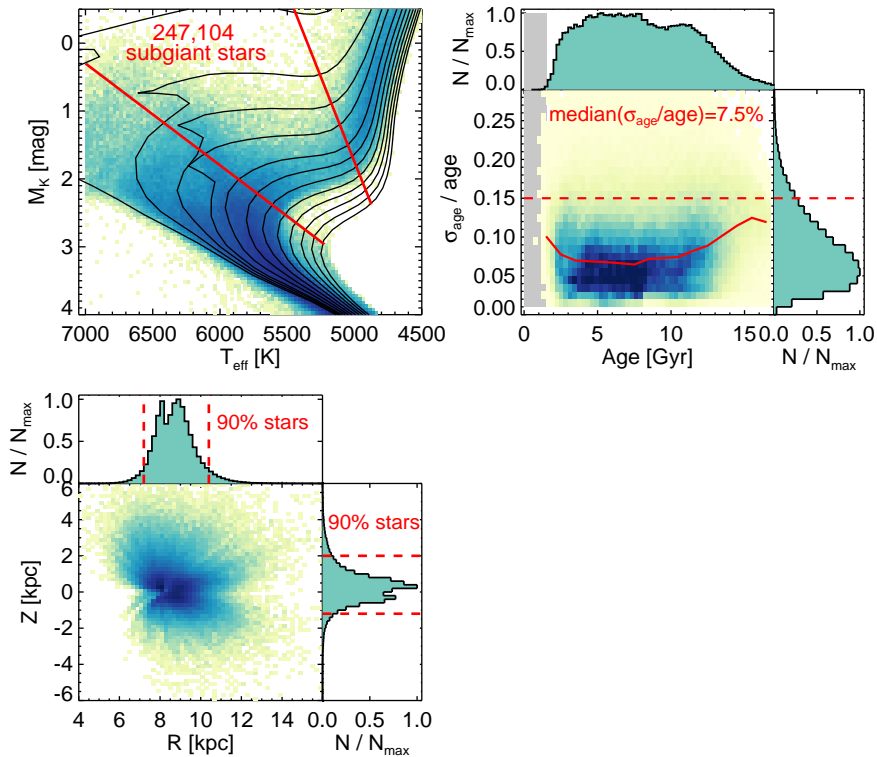}
\caption{\small {\bf The subgiant star sample with precise ages.} \emph{Top left}: Illustration of the subgiant selection in the \teff-\mk diagram, shown for the solar metallicity bin of $-0.1<{\rm [Fe/H]}<0.1$. The solid curves are isochrones from the Yonsei-Yale stellar evolution models \citep{Demarque2004} for solar metallicity ($\feh=0$, $\afe=0$) for ages of 1, 2, 4, 6, 8, 10, 12, 14, 16, 18, and 20\,Gyr, illustrating how stellar ages can be determined from the position in the \teff-\mk, if \feh is known. The two straight lines bracket the region within which we define our subgiant star sample. \emph{Top right}: Distribution in the relative age precision as a function of age: the mode of this precision distribution is at 6\%, the median at $\sim$7\%. Such high precision has never been reached for any large sample of stars before.  For the subsequent analysis we will only use stars with a relative age precision of $<$15\% (horizontal dashed line). \emph{Bottom left:} Spatial distribution of our subgiant sample stars in the $R$-$Z$ plane of Galactic cylindrical coordinates. The full extent of the sample in Galactocentric radius is $6\lesssim{R}\lesssim14$\,kpc and in distance from the Galactic mid-plane is $-5\lesssim{Z}\lesssim6$\,kpc. The bulk of the sample (90\%) covers $7.2\lesssim{R}\lesssim10.4$\,kpc, and  $-1.2\lesssim{Z}\lesssim2$\,kpc, as illustrated by the dashed lines.}
\label{fig:fig1}
\end{figure}

\textbf{Our Galaxy's stellar age-metallicity distribution}
The photospheric metallicity of any subgiant star of age $\tau$ reflects the element composition of the gas from which it formed at epoch $\tau$ ago. The overall distribution of these stellar metallicities at different epochs $p(\tau,\mathrm{[Fe/H]})$, thus encodes the chemical enrichment history of our Milky Way galaxy. The top panel of \textbf{Figure~2} presents this distribution for our new data. It shows that the age-metallicity distribution exhibits a number of prominent and distinct sequences, including at least two age-separated sequences with $\feh>-1$, as well as a sequence of exclusively old stars at low metallicity, $\feh<-1$. The density of $p(\tau,\mathrm{[Fe/H]})$ may change with stellar orbit or Galactocentric radius, where our sample covers (6--14\, kpc;  \textbf{Figure 1}). Yet, the ``morphology'' of the distribution varies only little, allowing us to focus mainly on the radially averaged distribution $p(\tau,\mathrm{[Fe/H]})$ here.

\begin{figure}[htp!]
\centering
\includegraphics[width=0.8\textwidth]{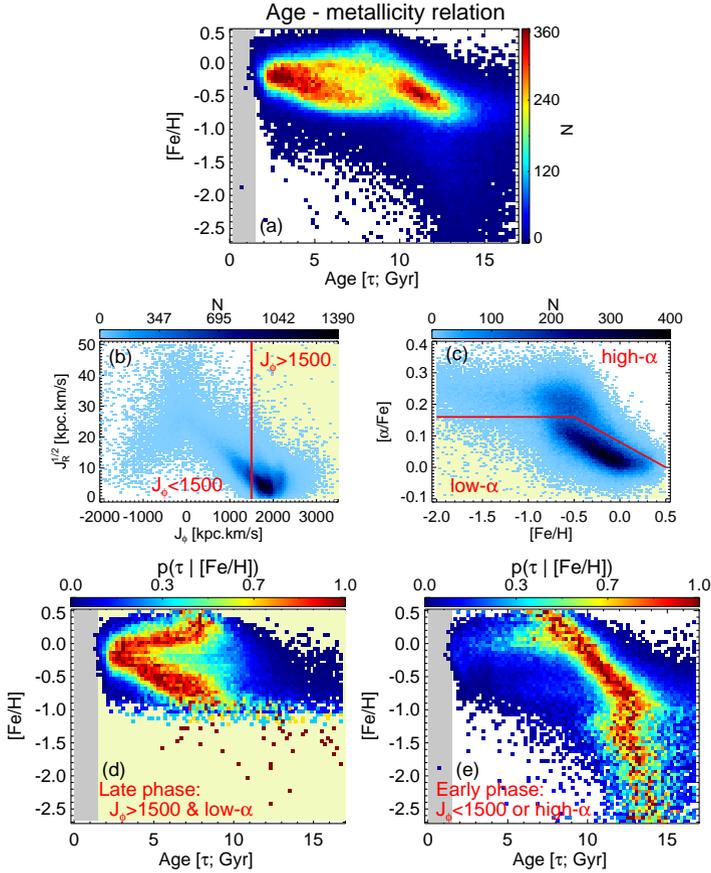}
\caption{\small  {\bf Stellar age-metallicity relation revealed by our subgiant star sample.} \emph{Panel (a)}: Stellar distribution in the age -- \feh plane for the whole subgiant star sample, color-coded by the stellar number density. \emph{Panel (b)}: Stellar density distribution in the plane of the azimuthal action $J_\phi$ (equivalent to angular momentum $L_Z$) versus radial action $J_R$. The vertical line delineates $J_\phi=1500$~kpc.km/s, which separates the sample into high angular momentum (yellow background) and low angular momentum regimes. \emph{Panel (c)}: Stellar density distribution in the [Fe/H]-[$\alpha$/Fe] plane. The red solid line separates the sample into high-$\alpha$ and low-$\alpha$ (yellow background) regimes. \emph{Panel (d)}: Probability distribution of stellar age $p(\tau\vert\feh)$, normalized to the peak value for each \feh, for stars with high angular momentum and low [$\alpha$/Fe] (yellow background regimes in the middle panels). \emph{Panel (e)}: Similar to the \emph{Panel (d)} but for stars with low angular momentum or high [$\alpha$/Fe]. The two regimes exhibit a sharp distinction at $\tau\simeq8$\,Gyr. Prominent structures are revealed for both regimes, such as the V-shape structure in the late phase ($Panel (d)$), the metal-poor ($\feh\lesssim-1$) `halo' and metal-rich ($\feh\gtrsim-1$) `disk' sequences in the early phase (bottom right panel). In the early phase, the two sequences merge at $\feh\simeq-1$, but the metal-rich sequence is older than the metal-poor sequence by $\sim$2\,Gyr at this metallicity, leading to a Z-shape structure in $(\tau\vert\feh)$.}
\label{fig:fig2}
\end{figure}

It turns out that the complexity of $p(\tau,\mathrm{[Fe/H]})$  (\textbf{Figure~2}, top panel) can be unravelled by dividing the sample into two sub-samples using stellar quantities that are neither $\tau$ nor \feh: the angular momentum $J_\phi$ (also denoted as $L_Z$) and the `$\alpha$-enhancement', [$\alpha$/Fe]. Extensive observations suggest that the majority of stars in the Milky Way formed from gradually enriched gas on high-angular momentum orbits, the extended (`thin') disk \cite{Frankel2018,Frankel+2020}, at high $J_\phi$ and low [$\alpha$/Fe]. It is also well established that the distribution of Galactic stars in the [$\alpha$/Fe]-[Fe/H] plane is bimodal: a high-$\alpha$ sequence reflecting rapid enrichment, and low-$\alpha$ sequence reflecting gradual enrichment, which suggests a natural way to divide any sample in the [$\alpha$/Fe]-[Fe/H] plane \cite{Hayden2015}. This inspired our approach to divide our sample into two, separating the dominant sample portion of gradually enriched disk stars with high angular momentum from the rest. Specifically, we used the cut
\begin{equation}
  \begin{cases}
    J_\phi > 1500\,{\mathrm {kpc.km/s}} \ \ \ \ \ {\tt and}\\
    \begin{cases}
    [\alpha/Fe] < 0.16, & \text{if $\feh < -0.5$}, \\
    [\alpha/Fe] < -0.16\feh + 0.08, & \text{if $\feh > -0.5$}, \\
    \end{cases}
  \end{cases}
\end{equation}
which is illustrated as a yellow shaded area in middle panels of \textbf{Figure~2}. The resulting subsamples in the $\tau$--\feh ~plane  are shown in the bottom two panels of \textbf{Figure~2}, where it is crucial to recall that the sample split involved neither of the quantities on the two axes, $\tau$ and \feh. Since we want to focus first on the Milky Way's elemental enrichment history, rather than its star-formation history, we normalize the distribution $p(\tau,\mathrm{[Fe/H]})$ at each \feh to yield $p(\tau\vert\feh)$, the age distribution at a given \feh. 

The bottom panels of \textbf{Figure~2} show that this cut in angular momentum and [$\alpha$/Fe] separates the Milky Way's enrichment history neatly into two distinct age regimes, with a rather sharp transition at $\tau\simeq8$~Gyr. We will therefore refer to these two portions, not clearly apparent in earlier data, as $p(\tau\vert\feh)_{\rm late}$ and $p(\tau\vert\feh)_{\rm early}$. The distribution of $p(\tau\vert\feh)_{\rm late}$ exhibits a V-shape that had not been seen that clearly before \cite{Feuillet2018}. This shape is presumably a consequence of the secular evolution of the quiescent disk -- the metal-rich ($\feh\gtrsim-0.1$) branch arises from stars that have migrated from the inner disk to near the Solar radius. The slope of that branch in $p(\tau\vert\feh)_{\rm late}$ then results from the (negative) radial metallicity gradient in the disk \citep{Xiang2015} and the fact that migrated more needed more time to do so, and are hence older. Analogously, we presume the lower branch of $p(\tau\vert\feh)_{\rm late}$ at $\feh\lesssim-0.1$ to arise from stars that were born further out and have migrated inward \citep{Wu2021}. A quantitative comparison with secular evolution models of the Galactic disc \citep{Frankel2018, Frankel+2020} is part of separate ongoing work. 

The older stars, reflected in $p(\tau\vert\feh)_{\rm early}$, exhibit two prominent sequences with distinct $\feh (\tau)$ relations. The stars with $-2.5<\feh<-1.0$ reflect the well-established the stellar halo population of our Milky Way, while the more metal-rich sequence ($\feh\gtrsim-1$) reflects the Milky Way's inner, high-$\alpha$ (`thick') disk \cite{Haywood2013}; this designation as an old disk component is also justified by the stars' angular momentum, as we will show below. 

This morphology of the old disk sequence in $p(\tau\vert\feh)_{\rm early}$ is the most striking feature in the bottom right panel of \textbf{Figure~2}: it reveals an exceptionally clear, continuous and tight age-metallicity relation from $\feh\lesssim-1$ 13\,Gyr ago all the way to $\feh=0.5$ $\sim$7\,Gyr ago. A simple model for $p(\tau\vert\feh)$ of this sequence (see online supplementary information) finds an intrinsic age dispersion of less than $0.82\pm0.01$\,Gyr at a given \feh across this 6\,Gyrs interval (see \textbf{Extended Figure 1}). Given the sequence's slope, this implies that the \feh dispersion at a given age is smaller than 0.22\,dex across the 1.5\,dex range in \feh.

Both the halo and old disk sequences extend to $\feh\simeq-1$. Whereas at that \feh value, the old disk sequence is $\sim$2\,Gyr older than the halo sequence, leading to a Z-shape structure in $p(\tau\vert\feh)_{\rm early}$. This offset is a second aspect of the distribution that has not been seen clearly before \citep{Montalban2021}. 

\textbf{Formation and enrichment of the Milky Way's old disk}
Tentative hints for some of these features in $p(\tau\vert\feh)$ have been seen in earlier work \cite{Haywood2013, Nissen2020} (see discussion in online supplementary information) but lacked the sample size or precision for definitive inferences about the Galactic formation history.
\textbf{Figure~2} reveals clearly that the old, high-$\alpha$ `thick' disk of our Milky Way started to form $\sim$13\,Gyr ago, which is only 0.8~Gyr after the Big Bang \cite{Planck2016}, and extended over 6~Gyrs, while the interstellar stellar medium (ISM) forming the stars was continually enriched by more than a dex, from \feh $\simeq-1$ to 0.5. The tightness of this \feh -age sequence implies that the ISM must have remained spatially mixed quite thoroughly during this entire period. Had there been any radial (or azimuthal) \feh variations (or gradients) in excess of 0.2\,dex in the star-forming ISM at any time, this would increase the resulting \feh -age scatter beyond what is seen. Such gradients, along with orbital migration, are the main reason that the later Galactic disk shows a considerably higher \feh dispersion at a given age \citep[\textbf{Figure~2} \& Ref.~][]{SchonrichBinney2009, Frankel2018}. 
 The results also show that the formation of Milky Way's old, $\alpha$-enhanced disk overlapped in time with the formation of the halo stars: the earliest disk stars are 1--2\,Gyr older than the major halo populations, at $\feh\simeq-1$ (see the Z-shape structure).  \\

\begin{figure}[htp!]
\centering
\includegraphics[width=0.8\textwidth]{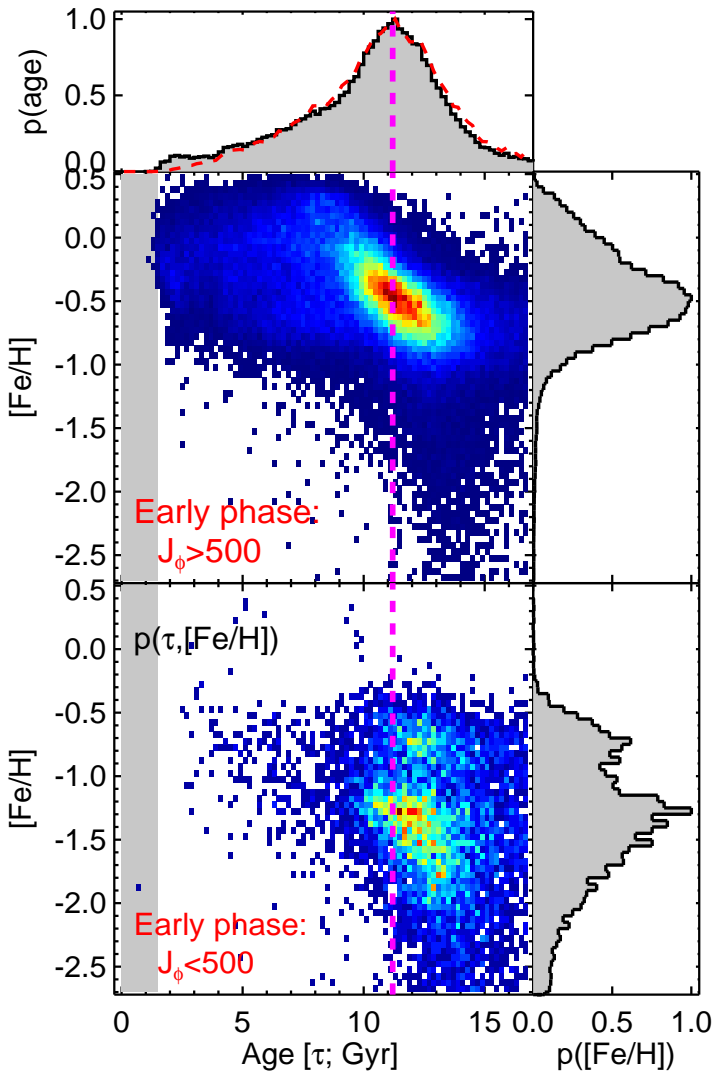}
\caption{\small {\bf Probability of stellar distribution in the $J_\phi$ versus \feh plane, $p(\tau, \feh)$, for stars formed in the early phase.} The stars formed in the early phase are divided into $J_\phi>500$\,kpc.km/s (upper) and $J_\phi<500$\,kpc.km/s (lower). The stellar distribution probability is normalized to the peak value so that the color from blue to red represents a value from 0 to unity. Note that this is different to $p(\tau\vert\feh)$ in \textbf{Figure~2}, which is normalized for each \feh. The histograms show the distribution integrated over \feh (top panel) or age (right panels). In the top panel, the age distribution $p(\tau)$ is a measure of the relative star-formation history. The dashed curve in red is for result after correcting for the volume selection effect. The vertical dashed line delineates a constant age of 11.2\,Gyr, when the star-formation rate reaches the maximum.}
\label{fig:fig3}
\end{figure}

In \textbf{Figure~3} we examine the $p(\tau\vert\feh)_{\rm early}$ distribution more closely by separating stars with at least modest angular momentum, $J_\phi > 500$\,kpc.km/s, from those stars on nearly radial or even retrograde orbits, $J_\phi < 500$\,kpc.km/s. This further sample differentiation by angular momentum leads again to two nearly disjoint $p(\tau\vert\feh)$ distributions. The first (\textbf{Figure~3}, upper panel), mostly $\feh>-1$, is dominated by the tight $p(\tau\vert\feh)$-sequence that we we have already attributed to the old disk. The second, predominately $\feh < -1.2$, reflects the halo. 

Note that the lower panel shows a distinct set of stars with $J_\phi < 500$\,kpc.km/s, whose $p(\tau, \feh)$ locus suggests that they are the oldest and most metal-poor part of the old disk sequence(see also \textbf{Extended Figure~2}). These stars suggest that some of the oldest members of the old disk sequence were present during an early merger event, by which they were ``splashed'' to low-angular-momentum orbits \cite{Bonaca2017, Belokurov2020}. This ancient merger event is presumably the merger with the Gaia-Enceladus satellite galaxy \cite{Helmi2018} (also known as Gaia Sausage \cite{Belokurov2018}), which has contributed most of the Milky Way's halo stars \cite{Di_Matteo2019, Helmi2020}. The fact that the splashed old disk stars with very little angular momentum are exclusively seen at $\tau\gtrsim11$\,Gyr constitutes strong evidence that the major merger process between the old disk and the Gaia-Enceladus satellite galaxy was largely completed 11\,Gyr ago. This epoch is 1\,Gyr earlier than previous estimates that had been based on the lower age limit of the halo stars, 10\,Gyr \cite{Helmi2018, Koppelman2018, Montalban2021}. 

\textbf{Figure~3} shows the volume-corrected two-dimensional distribution $p(\tau,\feh)$ (see online material for the correction of volume selection effect), rather than the $p(\tau \vert \feh)$ of \textbf{Figure~2}. This Figure reveals a remarkable new feature, namely that the star-formation rate of the old disk reached a prominent maximum at $\sim$11.2\,Gyr ago, apparently just when the merger with the Gaia-Enceladus was completed, and then continuously declined with time. The most obvious interpretation of this coincidence is that the perturbation from the Gaia-Sausage-Enceladus satellite galaxy greatly enhanced the star formation of the old disk.\\
  
To put our discovery into a bigger picture of galaxy formation and evolution, the multiple assembly phases are seen to be universal among present-day star forming galaxies. Using the IllustriesTNG simulation, Wang et al. \cite{Wang2022} showed that galaxy mergers and interactions have played a crucial role in inducing gas inflow, resulting in multiple star formation episodes, intermitted by quiescent phases. Observationally, the best testbed for this theoretical picture would be here at home within our Galaxy. Our study has demonstrated the power of such tests for galactic assembly and enrichment history in the full cosmic timeline, from the very early epoch ($\tau\simeq13\,Gyr$, or $z>10$) to the current time.

\bmhead{Acknowledgments}
The authors acknowledge Dandan Xu and Neige Frankel for helpful discussion, and Jan Rybizki for his kind help for using the Gaia mock catalogs. M.X. acknowledge partial support from the NSFC grant. 11833006 during his academic visit to NAOC during Nov. 2021 and Jan. 2022. This work has used data products from the Guoshoujing Telescope (LAMOST). LAMOST is a National Major Scientific Project built by the Chinese Academy of Sciences. Funding for the project has been provided by the National Development and Reform Commission. LAMOST is operated and managed by the National Astronomical Observatories, Chinese Academy of Sciences. This work has made use of data products from the European Space Agency (ESA) space mission Gaia. Gaia data are being processed by the Gaia Data Processing and Analysis Consortium (DPAC). Funding for the DPAC is provided by national institutions, in particular the institutions participating in the Gaia MultiLateral Agreement. The Gaia mission website is https://www.cosmos.esa.int/gaia. The Gaia archive website is https://archives.esac.esa.int/gaia. This publication has also used data products from the 2MASS, which is a joint project of the University of Massachusetts and the Infrared Processing and Analysis Center/California Institute of Technology, funded by the National Aeronautics and Space Administration and the National Science Foundation. 

\bmhead{Author contributions}
M. Xiang conducted the construction of the subgiant sample and the determination of stellar parameters and ages. M. Xiang and H.-W. Rix jointly executed the  data analysis and manuscript writing. 
\\

\clearpage
\noindent\textbf{\large Method}

\noindent{\bf Stellar labels from Spectroscopy:~}  Building this sample of subgiant stars with precise ages, abundances, and orbits requires a number of steps.
The first step is to derive stellar atmospheric parameters from the LAMOST DR7 spectra, which we did using the data-driven Payne (DD-Payne), verified in detail using analogous data from LAMOST DR5 \cite{Xiang2019}. This leads to a catalog of effective temperature \teff, surface gravity \logg,  microturbulent velocity $v_{mic}$, and elemental for 16 elements (C, N, O, Na, Mg, Al, Si, Ca, Ti, Cr, Mn, Fe, Co, Ni, Cu, Ba) for 7 million stars. We also derive an $\alpha$-element to iron abundance ratio [$\alpha$/Fe], which will serve in the age estimation to identify the right set of isochrones for each object. For a spectral S/N higher than 50, the typical measurement uncertainties are about 30\,K in \teff, and 0.05\,dex in the abundances we use here: [Fe/H] and [$\alpha$/Fe] \cite{Xiang2019}. 

\noindent{\bf Absolute Magnitude and Spectroscopic Parallax:~} Determining accurate and precise absolute magnitudes is crucial for age determination of subgiant stars (see top left panel of \textbf{Figure~1}). The Gaia astrometry provides high-precision parallax for stars within $\sim$2\,kpc, whereas for more distant stars the Gaia parallaxes have uncertainties in excess of 10\%. For these distant stars, spectroscopic estimates of absolute magnitude are needed to ensure precise age determination. 
We derive \mk, the absolute magnitude in 2MASS K band, from the LAMOST spectra and Gaia parallaxes, using a data-driven method based on neural network modelling (see Supplementary Information for details). \textbf{Extended Figure~3} illustrates that for LAMOST spectra with high signal-to-noise ratio ($S/N>80$), our spectroscopic \mk estimates are precise to better than 0.1\,mag at \feh =0 (and 0.15\,mag at \feh=$-1$). Furthermore, a comparison between spectrosopic \mk and astrometric \mk from Gaia parallaxes provides an efficient way of identifying unresolved binaries (\textbf{Extended Figure~3}) \cite{Xiang2019, Xiang2021}. For the subsequent modelling we combine these two approaches through a weighted mean algorithm
\begin{equation}
   \mk = \frac{\mk^{geom}/{\sigma_{geom}^2} +\mk^{spec}/{\sigma_{spec}^2}}        {{\sigma_{spec}^{-2}} + {\sigma_{geom}^{-2}} }.
\end{equation}

We are then in a position to select subgiant stars as lying between the two straight lines in the \teff-\mk diagram. As isochrones depend on \feh, this is done separately for each [Fe/H] bin, with the adopted slopes and intercepts for the boundary lines presented in \textbf{Extended Table~1}. As an example, the boundaries for stars with solar metallicity are shown in the top left panel of \textbf{Figure~1}. To ensure the boundaries vary smoothly with [Fe/H], we interpolate the slopes and intercepts listed the \textbf{Extended Table~1} to match the measured [Fe/H] for each star.  

\noindent{\bf Cleaning Sample Cuts:~} To have a subgiant star sample with high purity, we have applied cleaning criteria for discarding stars with bad data quality or stars that are possible contamination to the subgiant sample, this includes
\begin{itemize}
    \item We discard unresovled binaries which we identify via differences between their spectro-photometric parallax and their geometric parallax from Gaia, by requiring 
    \begin{equation}
        \frac{\varpi_{spec-photo} - \varpi_{geom}}{\sqrt{\sigma_{spec}^2 + \sigma_{geom}^2 }} < 2
    \end{equation} 
    Here $\varpi_{spec-photo}$ is the spectro-photometric parallax deduced from distance modulus using the spectroscopic \mk and 2MASS apparent magnitudes \citep{Skrutskie2006}.
    \item We discard stars with spurious Gaia astrometry using RUWE $>1.2$ or an astrometric fidelity $<$ 0.8 \cite{Rybizki2021}.
    \item We discard stars that show significant flux variability according to the variation amplitude of the Gaia magnitudes among different epoch, 
    \begin{equation*}
    \Delta_G = \frac{\sqrt{\rm {PHOT\_G\_N\_OBS}}}{\rm {PHOT\_G\_MEAN\_FLUX\_OVER\_ERROR}}
    \end{equation*}
    We calculate the ensemble median and dispersion of $\Delta_G$ as a function of G-band magnitude, and define any one star as a variable if 
    \begin{equation}
     \frac{\Delta_G - \overline{\Delta_G}}{\sigma(\Delta_G)} > 3 
    \end{equation} 
     Most of the variables eliminated by this criterion are found to be pre-main-sequence stars.
    \item We discard stars that are less luminous than the subgiant branch of a 20\,Gyr isochrone, the boundary of our isochrone grid. Such stars are mainly contaminations of either pre-main-sequence stars or main-sequence binary stars that are survived from the above criteria. 
    \item We discard all stars with \mk brighter than 0.5\,mag to avoid contamination from He-burning horizontal branch (HB) stars. This comes at a price: we eliminate essentially all stars younger than about 1.5\,Gyr. 
    \item We require all stars in our sample to have LAMOST spectral $S/N>20$, and with good DD-Payne fits, by requiring `qflag\_$\chi^2$ = good' \cite{Xiang2019}. We further restrict our stars to be $\teff<6800$\,K, where DD-Payne abundances are most robust.
\end{itemize}
After these cleaning cuts, the remaining sample contains 247,104 stars (\textbf{Figure~1}), presumed to be subgiants.

\noindent{\bf Age Estimates via Isochrones:~} The ages of the subgiant sample stars are determined by matching the astrometric parallax $\varpi$, spectroscopic stellar parameters {\teff,\mk, \feh, [$\alpha$/Fe]}, and Gaia and 2MASS photometry in $G, BP, RP, J, H, K$ bands, with the Yonsei-Yale (YY) stellar isochrones \cite{Yi2001, Demarque2004} with a Bayesian approach (see Supplementary Information for details), matched in both \feh \emph{and} [$\alpha$/Fe]. 
Note that in our Bayesian model, we have chosen not to impose a prior that all stars should be younger than the current knowledge on the age of the universe from CMB measurements of Planck (13.8\,Gyr) \cite{Planck2016}. This is for two main considerations. First, the upper limit of the stellar age is an independent examination of the age of the universe, whereas imposing age priors on the cosmological model inference might induce bias to the results. Second, imposing an upper age limit may increase complexity for statistics.  

To convert the Gaia parallax to absolute magnitudes, we also need to know the extinction. We therefore have determined the reddening and extinction for individual stars using intrinsic colors empirically inferred from their stellar parameters (see Supplementary Information for details). 

We have also tested the age estimation using other public isochrones, such as the MIST \citep{Dotter2016, Choi2016}, and find that in the case of solar $\alpha$-mixture, the age estimates based on YY and MIST show good consistency except for that the MIST isochrones predict 0.5\,Gyr older ages (\textbf{Extended Figure~4}). However, the $\alpha$-element enhancement, which is not avialble in the current public MIST isochrones, has a large impact on the age estimation, and ignoring the $\alpha$-element enhancement will lead to an overestimate of stellar age by up to 2\,Gyr for old stars (\textbf{Extended Figure~4}). Ages from the YY isochrones seem to be reasonable as they are consistent with the age of the universe at the old end (\textbf{Figure~2}).   

\noindent{\bf Orbital Actions:~} 
Using radial velocity from the LAMOST, proper motions from Gaia, and a combination of spectro-photometric distance and geometric distance (see Supplementary Information for details), we compute the orbital actions ($J_R$, $J_\phi$, $J_Z$) and angles of our sample stars using the $Galpy$ \cite{Bovy2015}, assuming the $MWPotential2014$ potential model. We assume the Sun is located at $R_\odot=8.178$\,kpc  \cite{Gravity2019}, and $Z_\odot = 10$\,pc above the disk mid-plane \citep{Xiang2018}. We assume LSR = 220\,km/s, and the solar motion with respect to the LSR, ($U_\odot$, $V_\odot$, $W_\odot$) = ($-7.01$\,km/s, 10.13\,km/s, 4.95\,km/s) \cite{Huang2015}.  

\noindent{\bf Accounting for selection effects:~} 
To verify that our findings are not caused by artefacts due to selection effects. We adopt two approaches to address this issue. First, we apply our target selection to the Gaia mock catalog of Rybizki et al. (2018)\cite{Rybizki2018}, and investigate the age--\feh relation (\textbf{Extended Figure~5}). Second, we directly correct for the volume selection function of our sample to account for the fact that, for a given light of sight, older subgiant stars probe to a smaller distance than the younger stars as the former are fainter. The age distribution of the thick disk stars after applying the selection function correction is illustrated in \textbf{Figure~3}. Eventually, we conclude that selection function has negligible impact on our conclusions (see Supplementary Information for more details). 

In addition, we have compared the stellar age-\feh relation from our sample with literature results for both stars \cite{Nissen2020} and Globular clusters \cite{Forbes2010,VandenBerg2013,Cohen2021} that have robust age estimates (see \textbf{Extended Figure~6}). The comparisons show qualitatively consistency, albeit the literature samples are too small to draw a clear picture of the assembly and enrichment history of our Galaxy (see Supplementary Information for a detailed discussion). 
\\

\bmhead{Data availability}
The Gaia eDR3 data is public available via https://www.cosmos.esa.int/web/gaia/earlydr3
The LAMOST DR7 spectra data set is public available via http://dr7.lamost.org. 

The subgiant star catalog generated and analysed in this study are provided as Supplementary Data, and it can also be reached via a temporary path https://keeper.mpdl.mpg.de/d/019ec71212934847bfed/. 

The Yonsi-Yale (YY) isochrones adopted for age determination in this work is public available via http://www.astro.yale.edu/demarque/yyiso.html. 

\bmhead{Code availability}
The stellar orbit computation tool $Galpy$ adopted in this work is public available via http://github.com/jobovy/galpy. 

The {\sc DD-Payne} code adopted for determining stellar labels, the neural network code for determining \mk from the LAMOST spectra, and the Bayesian code for stellar age estimation are currently not online public accessible, as they are planned to be applied to the upcoming LAMOST survey spectrum set. However, the codes can be shared by request for reasonable purpose.

\noindent\textbf{Additional information}

\noindent\textbf{Extended data} is available for this paper.

\noindent\textbf{Supplementary information} is available for this paper.

\bmhead{Competing interests}
The authors declare no competing interests.

\noindent\textbf{Reprints and permissions information} is available at http://www.nature.com/reprints.

\noindent\textbf{Correspondence and requests for materials} should be addressed to M. Xiang. 

\noindent\textbf{Author information} 

Maosheng Xiang, https:/orcid.org/0000-0002-5818-8769 

Hans-Walter Rix, https:/orcid.org/0000-0003-4996-9069

\newpage
\noindent\textbf{\large Extended Data} \\

\setcounter{figure}{0}
\renewcommand{\figurename}{Extended Figure}
\begin{figure}[htp]
\centering
\includegraphics[width=0.8\textwidth]{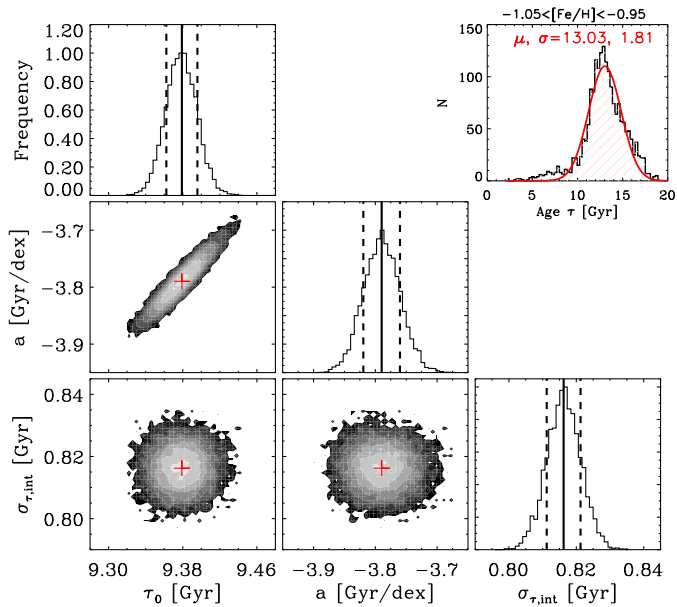}
\caption
{\small {\bf MCMC determination of the intrinsic scatter of the age distribution of old, high-$\alpha$ (`thick') disk sequence, $P(\tau\vert\feh)$, shown in panel $(e)$ of \textbf{Figure~2}}. The parameters shown are: $\sigma_{\tau,{\rm int}}$ -- the intrinsic age scatter, $\bar{\tau}_0$ -- the mean stellar age at solar metallicity ([Fe/H] = 0), and a -- the slope of mean age as a function of \feh. Specifically, we assume the age distribution for given \feh is $P(\tau, \delta\tau \vert \feh, \bar{\tau}_0, a, \sigma_{\tau,{\rm int}}) =  G\left(\tau-\bar{\tau}(\feh), \sqrt{\sigma_{\tau,{\rm int}}^2+\delta\tau^2}\right)$, where $G$ is the Gaussian function, $\delta\tau$ the measurement error of the age $\tau$, and $\bar{\tau}(\feh) = \bar{\tau}_0 + a\times\feh$ (see Supplementary Information for details).
Vertical solid and dashed lines indicate the mean and 1$\sigma$ values of the estimated parameters. 
The resultant upper limit of the intrinsic age scatter $\sigma_{\tau,{\rm int}}$ of the `thick' disk sequence is $\sim0.82\pm0.01$\,Gyr. This indicates that, at a constant age, the upper limit of the `thick' disk intrinsic \feh dispersion is 0.22\,dex. The upper-right corner shows the age distribution for stars formed in the early phase but with $-1.05<{\rm [Fe/H]}<-0.95$, $J_\phi>500$kpc.km/s -- presumably the oldest thick disk stars. A Gaussian fit to  the distribution (red curve) yields a mean age of 13\,Gyr. } 
\label{fig:fige1}
\end{figure}

\begin{figure}[htp!]
\centering
\includegraphics[width=0.6\textwidth]{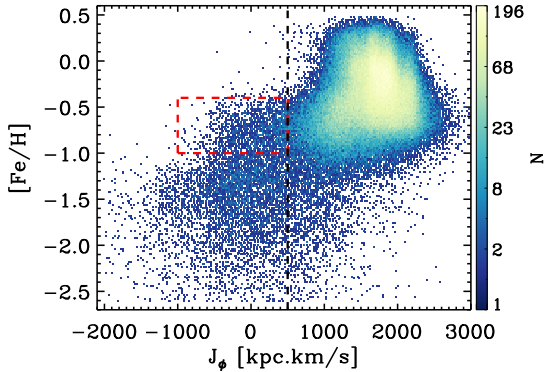}
\caption
{\small {\bf Stellar density distribution in the $J_\phi$ versus \feh plane}. The vertical line delineates a constant $J_\phi$ of 500\,kpc.km/s, which we adopt to separate the kinematic halo from the kinematic `thick' disk in \textbf{Figure~3}. There is a tail of low-angular-momentum stars ($J_\phi<500$\,kpc.km/s) in the metallicity range of $-1\lesssim\feh\lesssim-0.4$ (box delineated by red dashed lines), presumably the `splashed' thick disk stars due to the merger with the Gaia-Sausage-Enceladus satellite galaxy.} 
\label{fig:fige2}
\end{figure}

\begin{figure*}[htp!]
\centering
\includegraphics[width=1.0\textwidth]{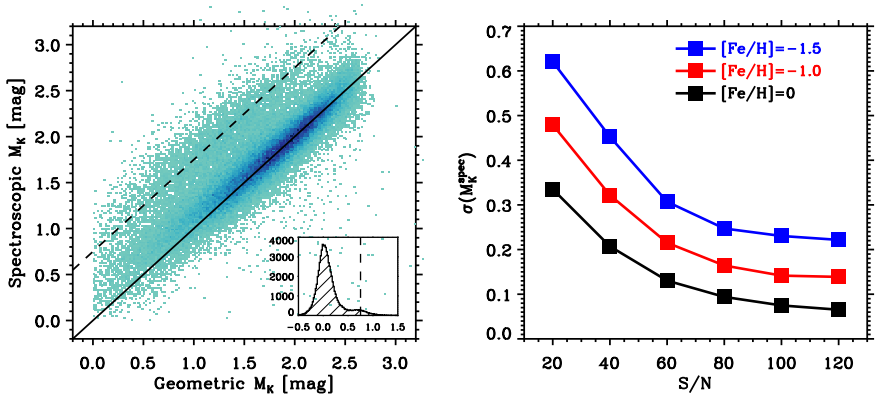}
\caption
{\small {\bf Validation of spectroscopic \mk estimates}. \emph{Left}: Spectroscopic \mk versus geometric \mk for a test set of stars with spectral $S/N>80$, $\sigma(M_K^{\rm geo})<0.2$\,mag. Colors indicate stellar number density. The stars with spectroscopic \mk much higher than geometric \mk are unresolved binaries, for which the geometric \mk are too luminous due to light contribution of the secondary. The solid line indicates the 1:1 line, and the dashed line indicates an offset of 0.75\,mag, which corresponds to the case of equal-mass binaries. The small window in the panel shows a histogram of the difference for spectroscopic \mk minus geometric \mk. \emph{Right}: uncertainty of the spectroscopic \mk estimates as a function of S/N, for subgiant stars of different metallicities. } 
\label{fig:fige3}
\end{figure*}

\begin{figure*}[htp]
\centering
\includegraphics[width=\textwidth]{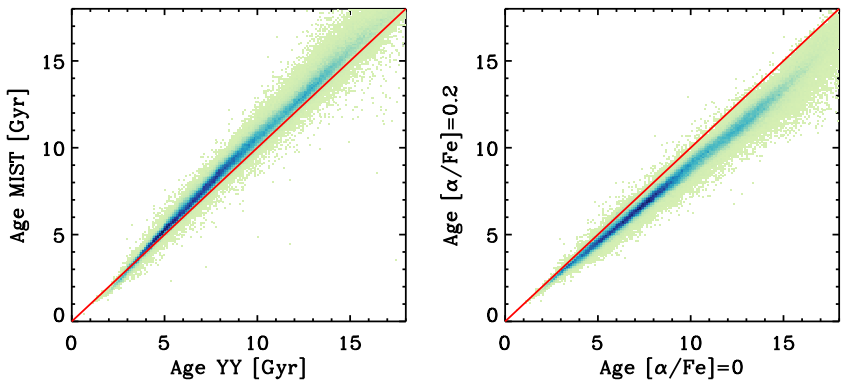}
\caption
{\small {\bf Illustration of age estimates from different isochrones}. \emph{Left}: comparison of age estimates from YY (X-axis) and MIST iscohrones (Y-axis), both with [$\alpha$/Fe]=0. MIST isochrones yield about 0.5\,Gyr older ages. Currently, MIST isochrones are publically available only with [$\alpha$/Fe] = 0, while YY isochrones with different $[\alpha/Fe]$  are available. \emph{Right}: Comparison of age estimates from YY isochrones with [$\alpha$/Fe] = 0 and with [$\alpha$/Fe] = 0.2. The 0.2\,dex $\alpha$-enhancement will alter the age estimates by 1--2\,Gyr, thus it is necessary to consider the this effect. We adopt the YY isochrones, and take the weighted-mean ages from isochrones with [$\alpha$/Fe] = 0, [$\alpha$/Fe] = 0.2, and [$\alpha$/Fe] = 0.4.} 
\label{fig:fige4}
\end{figure*}

\begin{figure*}[htp!]
\centering
\includegraphics[width=\textwidth]{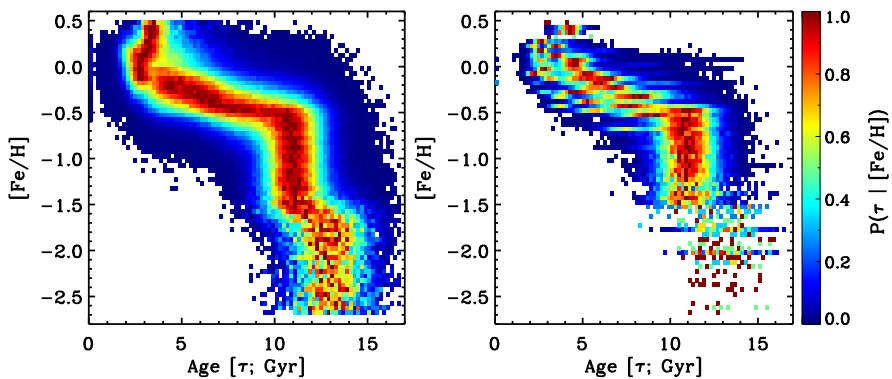}
\caption
{\small {\bf Examination of selection effect through Gaia Mock data}.  \emph{Left panel}: Age -- [Fe/H] relation for subgiant stars in the Gaia mock catalog of \cite{Rybizki2018}. The sample include 1,250,000 subgiant stars that in the same footprint and magnitude ranges as for the LAMOST. \emph{Right panel}: Same as the \emph{left} panel, but here for a subset of the Gaia mock subgiant stars that has comparable number of the LAMOST sample ($\sim$25,000 stars) randomly drawn from the sample shown in the \emph{left} panel. Compared to the \emph{left} panel, there are some artifacts for the younger populations ($\tau<9$\,Gyr) due to the smaller sample size, but this will not change the conclusion. } 
\label{fig:fige5} 
\end{figure*}

\begin{figure}[htp]
\centering
\includegraphics[width=0.8\textwidth]{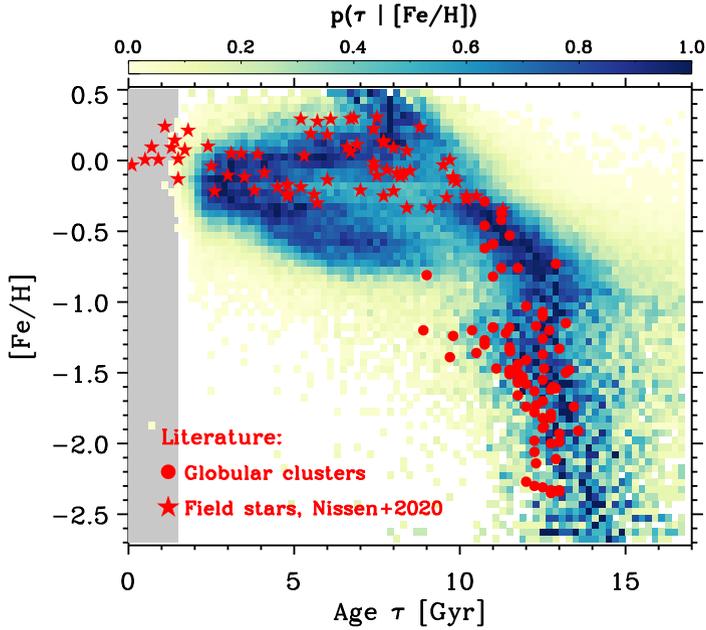}
\caption
{\small {\bf Comparison of the age-metallicity relation with literature}. The five-point stars in red represent field stars from Nissen et al. \cite{Nissen2020}, while the dots in red are globular clusters (GCs) compiled from Forbes et al. \cite{Forbes2010}, VandenBerg et al.\cite{VandenBerg2013}, and Cohen et al. \cite{Cohen2021}.  } 
\label{fig:fige6}
\end{figure}

\begin{table}
\centering
\caption{Slope and intercept of the linear functions for the upper and lower boundary of the subgiant star sample selection.}
\label{table:table1}
\begin{tabular}{ccccc}
\hline
[Fe/H] &  slope$_1$  & zpt$_1$ & slope$_2$ & zpt$_2$ \\
\hline
 $0.4$ & $-0.005$ & 26.00 & $-0.0014$ & 10.00 \\
 $0.2$ & $-0.005$ & 26.25 & $-0.0014$ & 10.10  \\
 $0.0$ & $-0.005$ & 26.50 & $-0.0014$ & 10.20 \\
 $-0.2$  & $-0.0045$ & 24.25 & $-0.0014$ & 10.30 \\
 $-0.4$ & $-0.004$ & 22.00 & $-0.0014$ & 10.50  \\
 $-0.6$  & $-0.004$  & 22.20 & $-0.0014$ & 10.70  \\
$-0.8$ & $-0.004$ & 22.60 & $-0.00125$ & 9.90  \\
$-1.0$  & $-0.004$  & 23.00 & $-0.00125$ & 10.00   \\
$-1.2$ & $-0.004$ & 23.20 & $-0.001$ & 8.65 \\
$-1.4$ & $-0.004$ & 23.40 & $-0.001$ & 8.70 \\  
$-1.6$ & $-0.004$ & 23.60 & $-0.001$ & 8.75 \\
$-1.8$ & $-0.004$ & 23.80 & $-0.001$ & 8.80 \\
$-2.0$  & $-0.004$ & 24.00 & $-0.001$ & 8.85 \\
$-2.2$  & $-0.004$ & 24.20 & $-0.001$ & 8.90 \\
$-2.5$  & $-0.004$ & 24.20  & $-0.001$ & 8.95 \\
 \hline
\end{tabular}
\begin{tablenotes}
      \small
      \item{The boundary of subgiant stars in the \teff--\mk diagram is \mk = slope$\times$\teff + zpt. The slopes and intercepts (`zpt') listed in the table are adopted as anchors for interpolation to match the measured [Fe/H] of each star.}.
\end{tablenotes}
\end{table} 


\begin{thebibliography}{99}
\bibitem[1]{Xiang2015} Xiang, M.-S. et al. The evolution of stellar metallicity gradients of the Milky Way disk from LSS-GAC main sequence turn-off stars: a two-phase disk formation history? Research in Astronomy and Astrophysics. \textbf{15}, 1209 (2015) 

\bibitem[2]{Bland-Hawthorn-Gerhard2016} Bland-Hawthorn, J. \&  Gerhard, O. The Galaxy in Context: Structural, Kinematic, and Integrated Properties. Annu. Rev. Astron. Astrophys. \textbf{54}, 529-596 (2016)  

\bibitem[3]{Spitoni2019} Spitoni, E., Silva Aguirre, V., Matteucci, F., Calura, F. \& Grisoni, V.  Galactic Archaeology with asteroseismic ages: Evidence for delayed gas infall in the formation of the Milky Way disc. Astron. Astrophys. \textbf{623}, A60 (2019)  

\bibitem[4]{Frankel2018} Frankel, N., Rix, H.-W., Ting, Y.-S., Ness, M. \& Hogg, D.~W. Measuring Radial Orbit Migration in the Galactic Disk. Astrophys. J. \textbf{865}, 96 (2018)  
 
\bibitem[5]{Feuillet2019} Feuillet, D.~K. et al. Spatial variations in the Milky Way disc metallicity-age relation. Mon. Not. R. Astron. Soc.  \textbf{489}, 1742-1752 (2019)

\bibitem[6]{Wu2021} Wu, Y.-Q. et al.  Age-metallicity dependent stellar kinematics of the Milky Way disc from LAMOST and Gaia. Mon. Not. R. Astron. Soc.  \textbf{501}, 4917-4934 (2021)

\bibitem[7]{Helmi2020} Helmi, A. Streams, Substructures, and the Early History of the Milky Way. Annu. Rev. Astron. Astrophys. \textbf{58}, 205-256 (2020)

\bibitem[8]{Hayden2015} Hayden, M.~R. et al. Chemical Cartography with APOGEE: Metallicity Distribution Functions and the Chemical Structure of the Milky Way Disk. Astrophys. J. \textbf{808}, 132 (2015)

\bibitem[9]{Bonaca2020} Bonaca, A. et al. Timing the Early Assembly of the Milky Way with the H3 Survey. Astrophys. J. \textbf{897}, L18 (2020)

\bibitem[10]{Belokurov2018} Belokurov, V., Erkal,  D., Evans, N.~W., Koposov, S.~E. \& Deason, A.~J. Co-formation of the disc and the stellar halo. Mon. Not. R. Astron. Soc.  \textbf{478}, 611-619 (2018)

\bibitem[11]{Helmi2018} Helmi, A., Babusiaux, C., Koppelman, H.~H., Massari, D., Veljanoski, J. \& Brown, A.~G.~A. The merger that led to the formation of the Milky Way's inner stellar halo and thick disk. Nature.  \textbf{563}, 85-88 (2018)

\bibitem[12]{Xiang2017} Xiang, M. et al. The Ages and Masses of a Million Galactic-disk Main-sequence Turnoff and Subgiant Stars from the LAMOST Galactic Spectroscopic Surveys. Astrophys. J. Suppl. Ser. \textbf{232}, 2 (2017)  

\bibitem[13]{Dotter2017} Dotter, A., Conroy, C., Cargile, P. \& Asplund, M. The Influence of Atomic Diffusion on Stellar Ages and Chemical Tagging. Astrophys. J. \textbf{840}, 99 (2017)

\bibitem[14]{Prusti2016} Gaia Collaboration. The Gaia mission. Astron. Astrophys. \textbf{595}, A1 (2016)

\bibitem[15]{Brown2021} Gaia Collaboration. The Influence of Atomic Diffusion on Stellar Ages and Chemical Tagging. Astron. Astrophys. \textbf{649}, A1 (2021)

\bibitem[16]{Cui2012} Cui, X.-Q. et al. The Large Sky Area Multi-Object Fiber Spectroscopic Telescope (LAMOST). Research in Astronomy and Astrophysics. \textbf{12}, 1197-1242 (2012)

\bibitem[17]{Zhao2012} Zhao, G., Zhao, Y.-H., Chu, Y.-Q., Jing, Y.-P. \& Deng, L.-C. LAMOST spectral survey -- An overview. Research in Astronomy and Astrophysics. \textbf{12}, 723-734 (2012)

\bibitem[18]{Demarque2004} Demarque, P., Woo, J.-H., Kim, Y.-C. \& Yi, S.~K. Y$^{2}$ Isochrones with an Improved Core Overshoot Treatment. Astrophys. J. Suppl. Ser. \textbf{155}, 667-674 (2004)

\bibitem[19]{Planck2016} Planck Collaboration. Planck 2015 results. XIII. Cosmological parameters. Astron. Astrophys. \textbf{594}, A13 (2016)

\bibitem[20]{SilvaAguirre2017} Silva Aguirre, V. et al.  Standing on the Shoulders of Dwarfs: the Kepler Asteroseismic LEGACY Sample. II. Radii, Masses, and Ages. Astrophys. J. \textbf{835}, 173 (2017)

\bibitem[21]{Montalban2021} Montalb{\'a}n, J. et al. Chronologically dating the early assembly of the Milky Way. Nature Astronomy. \textbf{5}, 640-647 (2021)

\bibitem[22]{Frankel+2020} Frankel, N., Sanders, J., Ting, Y.-S. \& Rix, H.-W. Keeping It Cool: Much Orbit Migration, yet Little Heating, in the Galactic Disk. Astrophys. J. \textbf{896},15 (2020)  

\bibitem[23]{Feuillet2018} Feuillet, D.~K. et al. Age-resolved chemistry of red giants in the solar neighbourhood. Mon. Not. R. Astron. Soc.  \textbf{477}, 2326-2348 (2018)

\bibitem[24]{SchonrichBinney2009} Sch{\"o}nrich, R. \& Binney, J. Chemical evolution with radial mixing. Mon. Not. R. Astron. Soc. \textbf{396}, 203-222 (2009)

\bibitem[25]{Haywood2013} Haywood, M.,  Di Matteo, P., Lehnert, M.~D., Katz,  D. \& G{\'o}mez, A. The age structure of stellar populations in the solar vicinity. Clues of a two-phase formation history of the Milky Way disk. Astron. Astrophys.  \textbf{560}, A109 (2013)

\bibitem[26]{Nissen2020} Nissen, P.~E.,  Christensen-Dalsgaard, J., Mosumgaard, J.~R.,  Silva Aguirre, V.,  Spitoni, E. \&  Verma, K. High-precision abundances of elements in solar-type stars. Evidence of two distinct sequences in abundance-age relations.  Astron. Astrophys. \textbf{640}, A81 (2020)

\bibitem[27]{Bonaca2017} Bonaca, A.,  Conroy, C., Wetzel, A., Hopkins, P.~F. \&  Kere{\v{s}}, D. Gaia Reveals a Metal-rich, in situ Component of the Local Stellar Halo. Astrophys. J. \textbf{845}, 101 (2017)

\bibitem[28]{Belokurov2020} Belokurov, V. et al. The biggest splash. Mon. Not. R. Astron. Soc. \textbf{494}, 3880-3898 (2020)

\bibitem[29]{Di_Matteo2019} Di Matteo, P. et al. The Milky Way has no in-situ halo other than the heated thick disc. Composition of the stellar halo and age-dating the last significant merger with Gaia DR2 and APOGEE. Astron. Astrophys. \textbf{632}, A4 (2019)

\bibitem[30]{Koppelman2018} Koppelman, H., Helmi, A. \& Veljanoski, J. One Large Blob and Many Streams Frosting the nearby Stellar Halo in Gaia DR2. Astrophys. J. \textbf{860}, L11 (2018)

\bibitem[31]{Wang2022} Wang, S. et al. From large-scale environment to CGM angular momentum to star-forming activities - I. Star-forming galaxies. Mon. Not. R. Astron. Soc. \textbf{509}, 3148-3162 (2022)\\
\end{thebibliography}

\begin{thebibliography}{99}
\bibitem[32]{Xiang2019} Xiang, M. et al. Abundance Estimates for 16 Elements in 6 Million Stars from LAMOST DR5 Low-Resolution Spectra. Astrophys. J. Suppl. Ser. \textbf{245}, 34 (2019)

\bibitem[33]{Xiang2021} Xiang, M. et al. Data-driven Spectroscopic Estimates of Absolute Magnitude, Distance, and Binarity: Method and Catalog of 16,002 O- and B-type Stars from LAMOST. Astrophys. J. Suppl. Ser. \textbf{253}, 22 (2021)

\bibitem[34]{Skrutskie2006} Skrutskie, M.~F. et al. The Two Micron All Sky Survey (2MASS).  Astron. J. \textbf{131}, 1163-1183 (2006)

\bibitem[35]{Rybizki2021} Rybizki, J. et al. A classifier for spurious astrometric solutions in Gaia EDR3. Mon. Not. R. Astron. Soc. \textbf{tmp}, 3298R, Preprint at https://arxiv.org/abs/2101.11641 (2021)

\bibitem[36]{Yi2001} Yi, S.~K. et al. Toward Better Age Estimates for Stellar Populations: The Y$^{2}$ Isochrones for Solar Mixture. Astrophys. J. Suppl. Ser. \textbf{136}, 417-437 (2001)

\bibitem[37]{Dotter2016} Dotter, A. MESA Isochrones and Stellar Tracks (MIST) 0: Methods for the Construction of Stellar Isochrones. Astrophys. J. Suppl. Ser. \textbf{222}, 8 (2016)

\bibitem[38]{Choi2016} Choi, J., Dotter, A., Conroy, C., Cantiello, M., Paxton, B. \& Johnson, B.~D. Mesa Isochrones and Stellar Tracks (MIST). I. Solar-scaled Models. Astrophys. J. \textbf{823}, 102 (2016)

\bibitem[39]{Bovy2015} Bovy, J. galpy: A python Library for Galactic Dynamics. Astrophys. J. Suppl. Ser. \textbf{216}, 29 (2015)

\bibitem[40]{Gravity2019} Gravity Collaboration. A geometric distance measurement to the Galactic center black hole with 0.3\% uncertainty. Astron. Astrophys. \textbf{625}, L10 (2019)

\bibitem[41]{Xiang2018} Xiang, M. et al. Stellar Mass Distribution and Star Formation History of the Galactic Disk Revealed by Mono-age Stellar Populations from LAMOST. Astrophys. J. Suppl. Ser. \textbf{237}, 33 (2018)  

\bibitem[42]{Huang2015} Huang, Y. et al. Determination of the local standard of rest using the LSS-GAC DR1. Mon. Not. R. Astron. Soc. \textbf{449}, 162-174 (2015)

\bibitem[43]{Rybizki2018} Rybizki, J., Demleitner, M.,  Fouesneau, M., Bailer-Jones, C., Rix, H.-W. \&  Andrae, R. A Gaia DR2 Mock Stellar Catalog. Publications of the Astronomical Society of the Pacific. \textbf{130}, 074101 (2018)  

\bibitem[45]{Forbes2010} Forbes, D.~A. \& Bridges, T. Accreted versus in situ Milky Way globular clusters. Mon. Not. R. Astron. Soc. \textbf{404}, 1203-1214 (2010)  

\bibitem[46]{VandenBerg2013} VandenBerg, D.~A., Brogaard, K., Leaman, R. \&  Casagrande, L. The Ages of 55 Globular Clusters as Determined Using an Improved $\Delta$V\^HB\_TO Method along with Color-Magnitude Diagram Constraints, and Their Implications for Broader Issues. Astrophys. J. \textbf{775}, 134 (2013)  

\bibitem[47]{Cohen2021} Cohen, R.~E., Bellini, A., Casagrande, L.,  Brown, T.~M., Correnti, M. \& Kalirai, J.~S. Relative Ages of Nine Inner Milky Way Globular Clusters from Proper Motion Cleaned Color-Magnitude Diagrams. Astron. J. \textbf{162}, 228 (2021)  
\end{thebibliography}
\end{document}